\documentclass[aps,prb,showpacs,twocolumn]{revtex4-2}
\def\today{16 December 2025}
\usepackage{graphicx}
\begin{document}

\title
{
\begin{minipage}[t]{7.0in}
\scriptsize
\begin{quote}
\leftline{{\it Phys. Rev. B}, in press.
}
\raggedleft {\rm arXiv:2507.13967}
\medskip
\end{quote}
\end{minipage}
\medskip
Excitonic Insulator and the Extended 
Falicov--Kimball Model Away from Half-Filling}

\author{D. I. Golosov}
\email{Denis.Golosov@biu.ac.il}
\affiliation{Department of Physics and the Resnick Institute, Bar-Ilan 
University, Ramat-Gan 52900, Israel.}

\date{\today}

\begin{abstract}
We consider an extended spinless 
Falicov--Kimball model  at an arbitrary doping level, focusing on the range of
parameter values where a uniform excitonic insulator is stabilised
at half-filling. In the two-dimensional case, we compare the properties of possible uniform phases and
construct the Hartree--Fock phase diagrams, which include sizeable
phase separation regions. It is seen that the excitonic insulator
can appear as a component phase in a mixed-phase state in a broad interval of
doping levels.
In addition, in a certain range of parameter values the {\it excitonic metal}
(doped excitonic insulator) is identified as the lowest-energy uniform phase.
We suggest that this phase, which is unstable with respect to phase separation, may be stabilised when the phase separation is suppressed by the long-range
Coulomb interaction. Overall, we find that excitonic correlations can affect
the behaviour of the system relatively far away from half-filling.

\typeout{polish abstract}
\end{abstract}
\maketitle

\section{INTRODUCTION}

Owing to its physical relevance and relative simplicity, the Falicov--Kimball
model (FKM) continues to attract much attention ever since its inception\cite{FK}
some 55 years ago. Theoretical investigations to date (involving a variety of
mean-field, numerical, and rigorous approaches) can be grouped in two broad
categories:
(i) Investigations of the model (with additional extensions -- Extended Falicov--Kimball model, EFKM) at half-filling $n=1$, when the number of carriers
equals the number of lattice sites\cite{Khomrev,Zlatic}. One of the most
prominent directions here is
related to the excitonic insulator (EI) -- a gapped phase of
condensed electron-hole pairs\cite{Khomskii76}, which is stabilised within the EFKM under
certain conditions (see Refs. \cite{Batista04,Farkasovsky08,prb2012} and many
others).
(ii) Studies of {\it doped} FKM with $n \neq 1$ (see, {\it e.g.}, Refs. \onlinecite{Gal,Freericks02,Farkasovsky05,Maska05}), and those of closely
related asymmetric Hubbard model\cite{Ueltschi,Farkasovsky08_2,Farkasovsky12}. Here, the ubiquitous finding is a strong
tendency toward phase separation/segregation\cite{Gal,Ueltschi,Farkasovsky12}, whereby, for example,
heavy electrons tend to congregate together in a part of the system only\cite{Freericks02,Farkasovsky08_2}.
It should be emphasised that all these studies without exception were
restricted to the values of parameters which do not allow for a stable uniform
EI phase in the half-filled case. This, in turn, implies that the EI is
altogether left out of any phase-separation scenario.

The objective of the present work is to begin filling this gap. While a complete
study would involve more sophisticated approaches and would also allow for
phases modulated by a non-zero wave vector, here we restrict ourselves to a
Hartree--Fock treatment and include only spatially-uniform phases
(along with phase separation between these). The key findings can be summarised as follows:
(i) Phase separation is expected to take place over a broad range of values
of parameters. In many cases, it includes a half-filed EI as one of the coexisting phases.
(ii) The mean-field equations allow for a uniform EI-type solution with
$n \neq 1$ in a broad range of concentrations $n$ around half-filling,
termed ``excitonic metal''. While it is thermodynamically unstable with
regard to phase separation, it presumably can be stabilised once the long-range
Coulomb repulsion is included in the model.

The overall conclusion is that excitonic physics, which in the context of FKM
was discussed in the half-filled case only, actually affects the behaviour
of the system over a relatively broad range of values of parameters, including
the doping level.

We construct phase diagrams of  doped EFKM at low
temperature, and discuss possible implications of our findings.

\section{THE MODEL AND MEAN-FIELD EQUATIONS}
\label{sec:mf}

We consider extended Falicov -- Kimball model (EFKM) with a Hamiltonian
\begin{eqnarray}
{\cal H}=&&-\frac{t}{2}\sum_{\langle i j \rangle} \left(c^\dagger_i c_j +
c^\dagger_j c_i \right) + E_d \sum_i d^\dagger_i d_i +U \sum_i c^\dagger_i
d^\dagger_i d_i c_i\, \nonumber\\
&&-\frac{t^\prime}{2}\sum_{\langle i j \rangle} \left(d^\dagger_i d_j+
d^\dagger_j d_i \right)\,.
\label{eq:FKM}
\end{eqnarray}
Here, the fermion operators $c_i$ and $d_i$ refer to electrons in the broad and
narrow bands (nearest-neighbour hopping parameters $t>|t^\prime|$), which
interact via on-site
repulsion $U$. We choose our units in such a way that both $t$ and the
period of the ($d$-dimensional hypercubic) lattice are equal to unity.
When the energy shift $E_d$ of the narrow band is equal to zero, the
Hamiltonian is identical to that of an asymmetric Hubbard model
(whereby one assigns opposite spins to the broad- and narrow-band carriers);
on the other hand, the case of ``pure'' Falicov--Kimball model
(as opposed to an {\it extended} one) is obtained from Eq. (\ref{eq:FKM})
in the limit
$t^\prime\rightarrow 0$. 

We perform Hartree--Fock decoupling in the Coulomb term,
\begin{eqnarray}
c^\dagger_id^\dagger_i d_i c_i \rightarrow && \,\,c^\dagger_ic_i n_d+d^\dagger_id_i n_c-n_cn_d - \nonumber \\
&& - d^\dagger_ic_i \Delta -  c^\dagger_id_i \Delta^* +|\Delta|^2\,,
\label{eq:decouple}
\end{eqnarray}
where $n_c$ and $n_d$ are carrier densities in the broad and narrow band,
and $\Delta=\langle c^\dagger_i d_i \rangle$ is the
off-diagonal (excitonic) average value; the latter is referred to as
induced hybridisation. We note that in the half-filled case, Hartree--Fock
approximation yields a remarkably good agreement with
quantum Monte Carlo simulations\cite{Batista04,Farkasovsky08}.

While spatially modulated mean field solutions of EFKM  may (and do) arise,
here we will be considering uniform solutions only, hence suppression of
the site index of the quantities $\Delta$ and $n_{c,d}$ (as well as of the
net density, $n=n_c+n_d$). Furthermore, we are primarily interested in the
region of parameters of Eq. (\ref{eq:FKM}), where the uniform EI
state (characterised by $\Delta\neq 0$) is stabilised at half-filling, $n=1$.
This implies\cite{Farkasovsky02,Batista04,Farkasovsky08,Czycholl08,prb2012} that both
$|E_d|$ and $|t^\prime|$ exceed certain critical values (with $t^\prime <0$ and
the critical value for $|t^\prime|$ being numerically small). The value of $U$
must be moderate in comparison to the width of the (broad) band,
$U\stackrel{<}{\sim} 2 d $.

The self-consistent mean-field equations for the {\it excitonic} phase are
derived
in a standard way ({\it e.g.,} by diagonalising the decoupled Hamiltonian)
and take form
\begin{eqnarray}
\Delta &=& \frac{1}{N} \sum_{\vec{k}} \frac{U \Delta\left(n^1_{\vec{k}}-n^2_{\vec{k}}\right)}{\sqrt{(\xi_{{\vec{k}}}
+t^\prime \epsilon_{\vec{k}})^2+4 |U\Delta|^2}} \,, \label{eq:delta}\\
n_c-n_d&=&\frac{1}{N}\sum_{\vec{k}} \frac{\left(\xi_{{\vec{k}}}+t^\prime \epsilon_{\vec{k}}\right)\left(n^1_{\vec{k}}-n^2_{\vec{k}}\right)}
{\sqrt{(\xi_{{\vec{k}}}
+t^\prime \epsilon_{\vec{k}})^2+4 |U\Delta|^2}}\,,
\label{eq:ndt}  
\end{eqnarray}
which is valid at all values of bandfilling $0<n<2$,
\begin{equation}
n=\frac{1}{N}\sum_{\vec{k}}\left(n^1_{\vec{k}}+n^2_{\vec{k}}\right)\,,\,\,\,
n^{1,2}_{\vec{k}}=\left({\rm e}^{\frac{\epsilon^{1,2}_{\vec{k}}-\mu}{T}}+1\right)^{-1}\,.
\label{eq:ntot}
\end{equation}
Here, $N$ is the total number of sites in the lattice, $\mu$ and $T$ are
chemical potential and temperature,
\begin{eqnarray}
  \epsilon^{1,2}_{\vec{k}}&=&\frac{1}{2}\left[E_d+Un+\left(1+t^\prime\right)\epsilon_k\right] \mp \nonumber \\
  &&\mp\frac{1}{2}\sqrt{(\xi_{{\vec{k}}}
  +t^\prime \epsilon_{\vec{k}})^2+4 |U\Delta|^2}
\label{eq:epsilon12}
\end{eqnarray}
are quasiparticle energies in the two new bands, $\epsilon_{\vec{k}}=-\cos k_x-
\cos k_y (- \cos k_z)$ is the tight-binding dispersion in two (three) dimensions,
and
\begin{equation}
  \xi_{\vec{k}}\equiv\xi(\epsilon_{\vec{k}})= E_d+U(n_c -n_d)-\epsilon_{\vec{k}}\,.
\label{eq:xi}
\end{equation}
Generally, the net energy
of the excitonic phase (``excitonic metal'', see Sec. \ref{sec:uniform} below) can be evaluated as [cf. Eq. (\ref{eq:decouple})]
\begin{equation}
  E_{EM}=\frac{1}{N} \sum_{\vec{k}} \left(\epsilon^1_{\vec{k}}n^1_{\vec{k}}+
  \epsilon^2_{\vec{k}}n^2_{\vec{k}}\right) +U\left(|\Delta|^2-n_cn_d\right)\,.
\label{eq:eem}
\end{equation}
At half-filling and at
$T \rightarrow 0$, the quantity $n^2_{\vec{k}}$ vanishes, while  $n^1_{\vec{k}} \equiv 1$ for
all values of $\vec{k}$; this corresponds to an EI, and Eq. (\ref{eq:eem}) then
yields $E_{EI}$, the energy of EI. We shall assume [without loss of generality -- see Eq. (\ref{eq:delta})]
that the quantity $\Delta$ is real and positive.

All other uniform mean field solutions do not have excitonic correlations ({\it i.e.,} $\Delta=0$) and fall into two categories.

First, there can be up to two {\it single-band} solutions, which are characterised
by one partially-filled band, the other band being either
completely filled
(at $n>1$) or empty (at $n<1$). These are prevalent for larger values of $U$ or $|E_d|$, or closer to the end points $n=0,2$.

At smaller $U$, one also finds {\it semimetal} solutions with two
partially-filled Hartree bands, centred around $Un_d$ (the broad band) and $E_d+Un_c=E_d+U(n-n_d)$ (the narrow band). Introducing $\lambda_c=\mu-U n_d$, we then find for the energy difference
$|t^\prime| \lambda_d$ between $\mu$ and the centre of the narrow band,
\begin{equation}
  |t^\prime| \lambda_d=\lambda_c-E_d+U(2n_d-n).
\label{eq:lambda}
\end{equation}
At $T\rightarrow 0$, the number of electrons in each band is found as
\begin{equation}
  n_d=\int_{-d}^{\lambda_d} \nu(\epsilon) d \epsilon\,,\,\,\,
  n-n_d=\int_{-d}^{\lambda_c} \nu(\epsilon) d \epsilon\,,
\end{equation}
where $\nu(\epsilon)$ is the tight-binding density of states in the broad band, and its argument is measured from the band centre.
Solving these three equations for the quantities $n_d$ and $\lambda_{c,d}$
yields up to three solutions, and we have to keep the one which is characterised
by the lowest energy,
\begin{equation}
  E=E_dn_d+Un_cn_d+\int_{-d}^{\lambda_c} \epsilon \nu(\epsilon) d\epsilon+
  |t^\prime| \int_{-d}^{\lambda_d} \epsilon \nu(\epsilon) d\epsilon\,.
\end{equation}

The rest of the paper is concerned with numerical analysis of these solutions
and of the ensuing phase diagrams in the interaction range $|t^\prime| \ll U/2d \stackrel{<}{\sim} 1$ (weak to moderately strong coupling). We shall be
interested in the
low-temperature limit, $T\rightarrow 0$ throughout.
While we consider the two-dimensional (2D) case only (aside from the Appendix),
it is expected that the mean-field analysis of a three-dimensional (3D) system
would
yield qualitatively similar results.  

\section{THE UNIFORM SOLUTIONS AND THEIR INSTABILITIES}
\label{sec:uniform}

We begin with the excitonic phase, which at half-filling, $n=1$,
corresponds to the well-known excitonic insulator (EI) state, characterised
by a non-zero value of the off-diagonal average $\Delta$ and a gap of the
order of $U\Delta$ in the electron spectrum. Properties of  EI phase and
its stability have been extensively addressed in the
literature\cite{Khomrev,Zlatic,prb2012}. Presently we
find, that once the EI solution is present at $n=1$, a similar solution to the
mean field equations, Eqs. (\ref{eq:delta}--\ref{eq:ndt}), persists also in a
certain range of doping values around $n=1$. The chemical potential then lies
within the upper or lower hybridised band, away from the excitonic band gap,
hence such a phase should be more properly called {\it excitonic metal} (EM).
This term, which was used previously in the context of doped Mott insulators\cite{Cyrot}, is now more commonly applied to the case of doped and/or otherwise
imperfect EI, with a non-zero density of states at the Fermi
level\cite{Veillette2001,Bi2021,Ghorai2023}.

\begin{figure}
\includegraphics{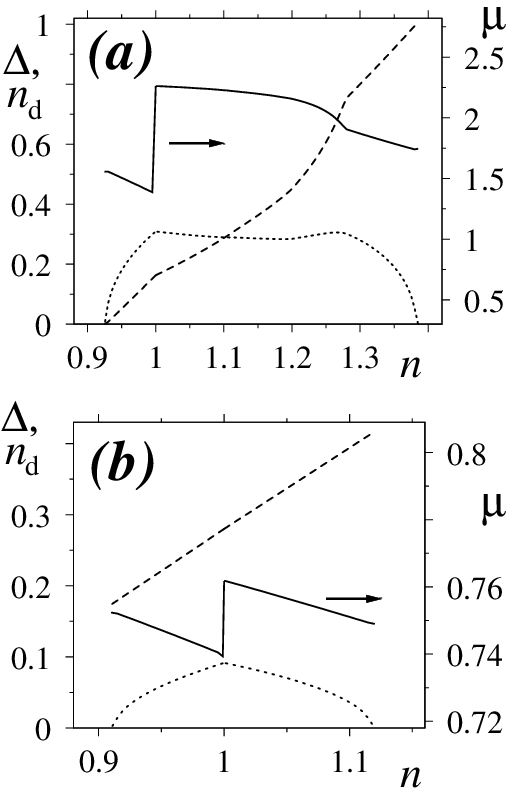}
\caption{\label{fig:EM} Excitonic metal solution for $U=2$, $E_d=0.4$, $t^\prime=-0.15$ {\it (a)} and for $U=0.5$, $E_d=0.4$, $t^\prime=-0.015$ {\it (b)}. Dashed and dotted lines show the dependence of $n_d$ and $\Delta$, respectively, on the carrier density $n$. Solid line (right scale) corresponds to the chemical potential $\mu$.}
\end{figure}

\begin{figure}
\includegraphics{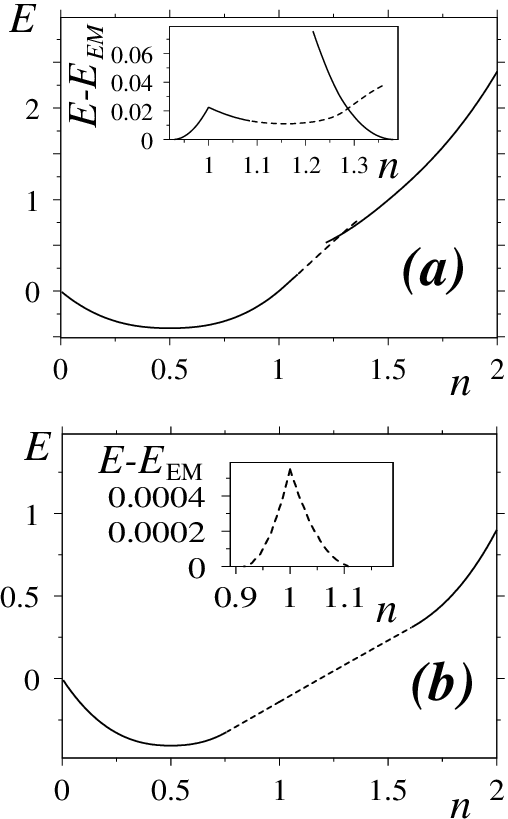}
\caption{\label{fig:esbsm}
Energies of single-band (solid line) and semimetal (dashed) solutions for $U=2$, $E_d=0.4$, $t^\prime=-0.15$ {\it (a)} and for $U=0.5$, $E_d=0.4$, $t^\prime=-0.015$ {\it (b)}. Insets show the energy differences between these solutions and the excitonic metal, the latter always corresponding to the lowest energy.}
\end{figure}

Fig. \ref{fig:EM} illustrates the typical behaviour of EM solution in the case of moderate {\it (a)}
or weak {\it (b)} Coulomb repulsion $U$. Numerical calculations are somewhat tedious, in
particular because even relatively small values of $t^\prime$ strongly affect
the EM bandstructure and may, {\it e.g.}, shift the minimum of the upper
hybridised band away from $\vec{k}=0$. For better convergence,  our
computations here and in Figs. \ref{fig:esbsm} and \ref{fig:phadiag} below
included a very
small temperature smearing (corresponding to $T=10^{-5}$) in the equations
(\ref{eq:delta}--\ref{eq:ndt}) for the EM phase away from half-filling; this
does not affect the
results, which therefore represent the $T\rightarrow 0$ case.

For $U=2$, the excitonic metal solution is found within the doping range $0.93 <n<1.38$
and with increasing $n$ the value of $n_d$ increases from zero to one.
Accordingly, at the two endpoints the excitonic solution merges with the two
different single-band phases. At these endpoints, the off-diagonal average
$\Delta$ shows square-root features; elsewhere, it has a maximum at $n=1$ and a smoother feature near $n=1.3$, the latter reflecting a feature in the
bandstructure. In the weak-interaction regime of $U=0.5$, the excitonic metal
solution arises for $0.91<n<1.12$, merging with the semimetal solution at the endpoints.

Importantly, everywhere away from half-filling the compressibility
$\partial \mu/\partial n$ is negative (see Appendix), signalling an
instability with respect
to an inhomogeneity formation and ultimately to phase separation. The dependence
$\mu(n)$ shows a jump at $n=1$. This point, where the value of $\mu$ lies
within the gap between the lower (filled) and the upper (empty) bands [see Eq.
  (\ref{eq:epsilon12})], has to be considered separately.

\begin{figure}
\includegraphics{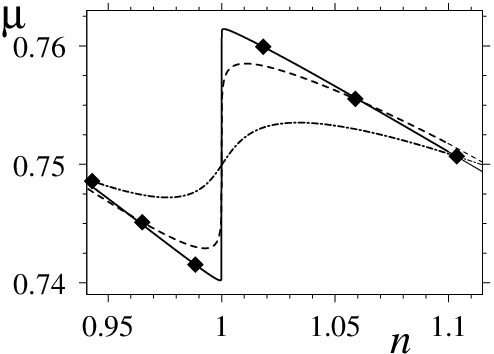}
\caption{\label{fig:halffilled} Chemical potential near half filling in the
  excitonic metal phase for $U=0.5$, $E_d=0.4$, and $t^\prime=-0.015$.
  Solid, dashed,
  and dashed-dotted lines corresponds to $T=10^{-4}$, $T=10^{-3}$, and $T=3 \cdot 10^{-3}$ in the Fermi distributions in Eqs. (\ref{eq:delta}--\ref{eq:ntot}).
  The diamonds correspond to respective chemical potential values crossing out
  of the hybridisation gap, whose width is about $0.02$.
  The outer pair of diamonds refer to $T=3 \cdot 10^{-3}$, and the middle one --
  to $T=10^{-3}$.}
\end{figure}

In this situation, the value of $\mu$ at $T=0$ can be defined only as the
$T\rightarrow 0$ limit of chemical potential at a finite $T$. In the latter
case, the value of $\mu$ is fixed owing to the smearing of the Fermi
distribution, which in turn gives rise to a non-degenerate gas of holes
(electrons) in the valence (conduction) band. Typical evolution of $\mu(n)$
with decreasing temperature\cite{distr} is shown in Fig. \ref{fig:halffilled}. 

We observe that at a finite temperature, $\mu(n)$ has a minimum below the point $n=1$ and a maximum above it. Numerical data show that when the chemical
potential attains its maximal (minimal) value, it lies within the energy
gap and the energy difference between $\mu$ and the bottom of the conduction band (the top of the valence band) is of the order of $T$. This entails
two conclusions regarding the low-temperature limit.

First, the compressibility at $n=1$ stays positive, increasing as the
temperature decreases (and ultimately diverging at $T\rightarrow 0$).
Thus, while excitonic metal at $n \neq 1$ is thermodynamically unstable, no
such instability is found for an excitonic insulator at $n=1$. This agrees
with the literature, confirming the stability of EI in the suitable
range of EFKM parameter values.  

Second, as long as the chemical potential lies within the gap
(and the distance from the gap edges is large in comparison with $T$),
the compressibility is positive. At $T\rightarrow 0$, this translates into the
following conclusion, which we will use in Sec. \ref{sec:phasep} below:
When the  value of $\mu$ is externally fixed and lies anywhere within the
gap, the EI phase (with $n \rightarrow 1$ at $T\rightarrow 0$) remains stable.

\begin{figure}
\includegraphics{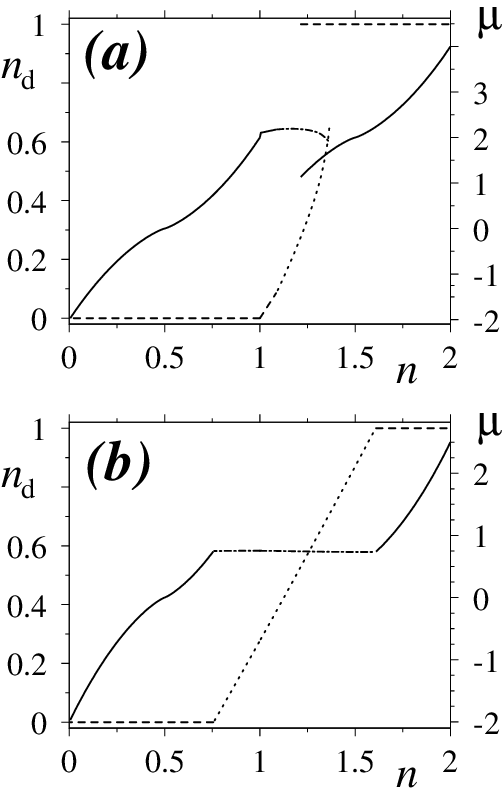}
\caption{\label{fig:sbsm}
  Single-band and semimetal solutions for $U=2$, $E_d=0.4$, $t^\prime=-0.15$ {\it (a)} and for $U=0.5$, $E_d=0.4$, $t^\prime=-0.015$ {\it (b)}. At a given value of carrier density $n$, dashed and dotted lines show the value of $n_d$ (left scale) for the lowest-energy single-band and semimetal solutions,
  respectively. Solid and dashed dotted lines represent the corresponding values of chemical potential $\mu$ (right scale).  }
\end{figure}

We now turn to the $\Delta=0$ solutions mentioned in Sec. \ref{sec:mf} above.
The behaviour of $n_d(n)$ and $\mu(n)$ for single-band
and semimetal solutions is illustrated in Fig. \ref{fig:sbsm}. We recall that
when the value of $U$ is sufficiently large, there can be up to three
different semimetal solutions for a given value of $n$, and (at still
larger $U$) up to two single-band solutions; we are always interested in the
lowest-energy solutions of both types. 

While in the weak coupling case (see Fig.  \ref{fig:sbsm} {\it b}) the
sequential single-band and semimetal solutions evolve continuously from $n=0$
to $n=2$, in the larger-$U$ case of  Fig.  \ref{fig:sbsm} {\it a} we find a
discontinuity. The latter reflects the presence of multiple semimetal solutions
in the region around $n=1.2$.

Single-band solutions, which are characterised by  $n_d=0$,  $n_d=1$,
$n_d=n-1$, or $n_d=n$, always have positive compressibility. This is not the
case for the
semimetal solutions, which in both cases shown in Fig. \ref{fig:sbsm}
($U=2$ and $U=0.5$) have negative $\partial \mu/\partial n$ for certain values
of $n$ above half-filling. Indeed, the compressibility of semi-metallic phase is given by
\begin{equation}
  \frac{\partial \mu}{\partial n}=\frac{U^2 \nu(\lambda_c) \nu(\lambda_d) -
    |t^\prime|}{2U \nu(\lambda_c) \nu(\lambda_d) -\nu(\lambda_d)-  |t^\prime|
    \nu(\lambda_c)}\,,
  \label{eq:smcompre}
\end{equation}
    [see Eq. (\ref{eq:lambda})]; since $\nu(\epsilon)$ diverges at $\epsilon \rightarrow 0$ and
    equals $1/2\pi$ at the band edge, this can change sign at $U< \pi$ (in the case of small $|t^\prime|$).
    
The energies of these single-band and semimetal solutions are plotted in Fig.
\ref{fig:esbsm}. Importantly, whenever the excitonic metal solution is present,
its energy $E_{EM}(n)$ [see Eq. (\ref{eq:eem})] is lower than that of other
uniform solutions, as shown in the insets in Fig. \ref{fig:esbsm}. This is
the typical situation, although a different behaviour can be found when all
three of
$U$, $E_d$, and $n$ take larger values (see below, Sec. \ref{sec:phadiag}).
The peak at $n=1$ (see insets in Fig. \ref{fig:esbsm} {\it a,b}) is due to a
sharp minimum of
$E_{EM}(n)$ at half-filling.

\begin{figure}
\includegraphics{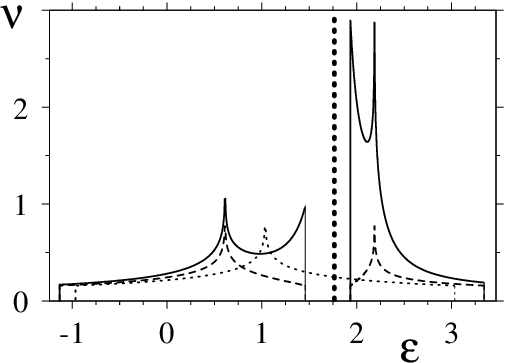}
\caption{\label{fig:dos} Typical energy dependence of the quasiparticle density of states (DOS) for an excitonic metal solution (solid lines). Dashed lines
  show the contributions of broad-band electrons to the net DOS, whereas the
  dotted lines correspond to the DOS in the absence of hybridisation
  (obtained by formally setting $\Delta=0$; bold dotted line represents the localised band).
  The data correspond to $n=1.2$, $U=2$, $E_d=0.4$. We present results obtained for $t^\prime \rightarrow 0$, as corrections due to a finite value of
  $t^\prime$ are not significant for our purposes here.}
\end{figure}

Thus,  the energy can be gained by opening the excitonic gap even away from
half-filling, with the chemical potential lying within one of the hybridised
bands.
This is because unlike in a conventional low-energy, longwavelength scenario,
excitonic pairing in the EFKM involves short-range correlations, hence
a restructuring of the spectrum over the entire bandwidth. The latter is illustrated in Fig. \ref{fig:dos}, where the dotted lines show the density of
states (DOS) without hybridisation: broad conduction band and a delta-functional
feature (bold dotted line) for the narrow band in the small-$t^\prime$ limit.
Solid lines show the DOS in the two hybridised bands
[see Eq. (\ref{eq:epsilon12})]. These are shifted away from the original position of the narrow band, giving rise
to a gap and also increasing the overall bandwidth. The maxima of the DOS
(usual logarithmic feature in 2D) are also moved away from the gap. The dashed
lines
denote the contribution of broad-band electrons to the DOS; we see that the
localised electrons not only contribute the cusps adjacent to the gap, but
also affect the DOS over the entire band. While the energy gain associated with
opening the gap would be maximal at half-filling, in the case of
Fig. \ref{fig:dos} the carrier density $n$ equals $1.2$, and the chemical
potential value for a hybridised solution is $\mu \approx 2.02$. Owing to
the overall redistribution of the DOS, the energy of the Fermi sea is still
lowered upon the gap opening.

We recall that while negative compressibility does indicate an instability
with respect to phase separation, the former is far from being a necessary
condition for the latter. Regardless of the sign of $\partial \mu/\partial n$
for single-band and semimetal solutions in the region where the lower-energy
excitonic
solution is present, the system always undergoes phase separation throughout
this region (excepting the point $n=1$). This is because the lowest-energy
(excitonic) solution has negative compressibility. In Secs. \ref{sec:phasep}, \ref{sec:phadiag}
we will see that this tendency towards phase separation actually extends well
beyond the range of densities of the excitonic solution.


We remark that analysis of a ``pure'' Falicov -- Kimball model
[Eq. (\ref{eq:FKM}) with $t^\prime=0$] yields qualitatively similar behaviour
of mean-field solutions, although quantitative changes associated with the
non-zero $t^\prime$ are in some cases appreciable. We include a finite
$t^\prime <0$ in our analysis due to the  peculiar features
of the $t^\prime=0$ case. These
include an instability of excitonic insulator at $n=1$
(for all values of $U$ and $E_d$), as clarified in
Refs. \cite{Farkasovsky08,prb2012}. For the numerical data shown in the plots, we chose the values of
$-t^\prime$ well above the respective critical values required to stabilise the homogeneous
excitonic insulator state at half-filling; these critical values correspond
to a second-order transition into a spatially modulated phase\cite{Farkasovsky08,prb2012}. 

It is well-known that at $n=1$ the excitonic insulator also shows an
instability of another kind,
which arises in the immediate vicinity of  the symmetric point
$E_d=0$ [for a broad range of values of $t^\prime$ in Eq.  (\ref{eq:FKM})].
There, previous work\cite{Batista04,Farkasovsky08,Czycholl08}
reports a charge/orbital-ordered state, emerging via a first-order
transition. Since presently we study uniform phases only,
this ordering is beyond the scope
of our approach, and therefore we do not detect any sign of the
associated instability. 

Finally, we note that in the Falicov--Kimball model the stability of
excitonic insulator at $n=1$ (away from the $E_d=0$ point) can also be restored
if, instead of the hopping $t^\prime$ in the narrow band,
bare hybridisation is added to the model\cite{Czycholl99,prb2012}. This corresponds to replacing the
last term in Eq. (\ref{eq:FKM}) with, {\it e.g.},

\begin{equation}
V_0 \sum_i c^\dagger_i d_i -\frac{V_1}{2}\sum_{\langle i j \rangle}\left( 
c^\dagger _i d_j + c^\dagger_j d_i \right)+ {\rm H. c.}\,,
\label{eq:Vpert}
\end{equation}

where the values of $V_0$ and/or $V_1$ exceed certain numerically small
critical values. While this leads to numerical changes, we tentatively
verified that at
the qualitative level, the conclusions of this section (that the excitonic
metal state has a negative compressibility and also, typically, the
lowest energy) remain valid. The important difference is that in the presence
of a bare hybridisation one has to distinguish not between
non-hybridised ($\Delta =0$) and hybridised (excitonic metal or insulator) solutions as above, but rather between
phases with small and
large $\Delta$. In the former case, $\Delta$ vanishes in the
limit $V_{0,1} \rightarrow 0$, as the corresponding solution evolves into either single-band or semimetal one. This does not happen in the case of large-$\Delta$ solutions, corresponding to EM or EI (notwithstanding
an instability at small $V_{0,1}$, which is similar to the instability at small $t^\prime$). Throughout
the rest of the paper, we will be including the effects of $t^\prime$ only.

\section{PHASE SEPARATION}
\label{sec:phasep}

In the context of FKM and the asymmetric Hubbard model,
the ubiquitous phenomenon of phase separation attracts much attention both in
the half-filled\cite{Kennedy98,Farkasovsky01} and doped ($n \neq 1$, Refs. \onlinecite{Freericks02,Ueltschi,Farkasovsky08_2,Maska05}) cases. However, the available
studies of the doped case treat the situation away from the region of
parameter values where the uniform EI state is stabilised for $n=1$. As
explained above, in terms of our Hamiltonian, Eq. (\ref{eq:FKM}), the latter
region corresponds to finite non-zero values of {\it both} $E_d$ and (negative) $t^\prime$. In this case the EI, being the lowest-energy state for $n=1$,
can become one of the two component phases in a phase-separated system.

First, let us turn to a generic situation of equilibrium between two
phases A and B, with respective energies $E_{A,B}(n)$ and chemical potentials
$\mu_{A,B}(n)$, which depend on the density $n$. 
The equilibrium condition involves
two equations for chemical potentials and Gibbs free energies:
\begin{eqnarray}
  \mu_A(n_A)&=& \mu_B (n_B) \,,
\label{eq:phasep1} \\
  E_A(n_A)-\mu_A(n_A) n_A &=& E_B(n_B) - \mu_B (n_B) n_B\, ,
  \label{eq:phasep2}
\end{eqnarray}
which determine the two values  $n_{A,B}$ of density in the respective
regions of phases A and B.

The preceding description refers to the situation when  phases A and B are both gapless, and
needs to be modified in the case of phase equilibrium between a gapless phase A and the
(gapped) half-filled EI (we call this phase separation of the first kind, PS1).
Chemical potential anywhere in the system will then be given by $\mu_A(n_A)$.
As explained in the previous section, the EI will remain stable (with $n_{EI}=1$ at $T\rightarrow0$) as long as $\mu$ lies between the top of the valence band
and the bottom of the conduction band:
\begin{equation}
  \epsilon^1_{max} < \mu_A(n_A) < \epsilon^2_{min}
  \label{eq:gap}
\end{equation}
[See Eq. (\ref{eq:epsilon12})]. The value of $n_A$ is then found from the equation for
Gibbs free energies, which takes form
\begin{equation}
F_A(n_A) \equiv E_A(n_A)-E_{EI}-(n_A-1) \mu_A(n_A) =0
\label{eq:F}
\end{equation}
[see Eq. (\ref{eq:eem})].
Conditions (\ref{eq:gap}--\ref{eq:F}) take place of Eqs. (\ref{eq:phasep1}--\ref{eq:phasep2}). Since presently we consider uniform phases only,
the phase A must be either single-band or semimetallic one. While this may
leave out some possible scenarios, we do not expect this condition to be too
restrictive: in principle, a (presumably gapped) homogeneous phase with
spatial modulation may become stabilised near a commensurate value of
density away from half-filling, yet this would require a large value of $U$.
Even then, it might prove impossible to bring it in equilibrium with the
half-filled EI.

We note that the
question of EI stability at half-filling has been addressed by a variety of
methods (see, {\it e.g.},
Refs. \onlinecite{Batista04,Farkasovsky08,Czycholl08,prb2012}), also beyond the simple
mean-field approximation. Within the range of parameter
values where the uniform EI is stable at $n=1$,  this half-filled EI will
therefore
be stable also in a phase-separated state
away from the (overall) half-filling, provided that the EI is in the
thermodynamic equilibrium [see Eqs. (\ref{eq:gap}--\ref{eq:F})] with the
other constituent phase.

There is an additional phase separation scenario (denoted PS2) which must be
taken into account, in particular  whenever the half-filled EI solution
disappears at large $|E_d|$, and/or for
at smaller $U$ ($U \stackrel{<}{\sim} 0.75$). This corresponds
to a phase separation into two different non-excitonic phases [A and B in Eqs.
  (\ref{eq:phasep1}--\ref{eq:phasep2})], which can be either single-band
with partially filled broad (SB1) or
narrow (SB2) band, or semimetal. As usual, we can use Eq. (\ref{eq:phasep1}) to express $n_B$ as a function of $n_A$, and the phase equilibrium takes place at
\begin{eqnarray}
  F_{AB}(n_A) \equiv &&E_A(n_A)-E_{B}\left(n_B(n_A)\right)- \nonumber \\
  &&-\left[n_A-n_B(n_A)\right] \mu_A(n_A) = 0.
\label{eq:FAB}
\end{eqnarray}
At smaller $U$, this can preempt the EI -- single-band PS1 phase separation and may also render the EI state at $n=1$ unstable (basically,
the single-band state that would have been in thermodynamic equilibrium with the EI becomes unstable with respect to this second type of phase separation). We will now follow these two phase separation scenarios in a typical situation
(the details may vary, depending on  parameter values).

Let us  imagine that we start with two fully occupied bands and gradually lower the
concentration $n$. Provided that the value of $E_d$ is less than half-bandwidth [more precisely, $E_d<2-2|t^\prime|$ in Eq.
  (\ref{eq:FKM})], this leads to a depletion of carriers in the broad band and gives rise to a
Fermi surface: the system is in a uniform SB1 phase, denoted phase $A$.
As the value of $n$ decreases further, the value of the chemical potential $\mu_A$ crosses
(from above) inside the range, corresponding to the spectral gap of a half-filled EI, Eq.
(\ref{eq:gap}), with the value of $F_A(E)$, Eq. (\ref{eq:F}),  being negative.
Also, as
we lower the carrier density from the $n=2$ endpoint, we eventually arrive at a point where a
solution to Eq. (\ref{eq:phasep1}) appears [see the two solid lines in Figs.
\ref{fig:sbsm} (a,b)], with phase $B$ being another single-band phase
(or possibly a semimetallic one). Initially, the value of
$F_{AB}$, Eq. (\ref{eq:FAB}), will also be negative.      

With lowering $n$ further, the stability of the uniform phase $A$ is lost
at a point $n=n_A^*$
where either $F_A$ or $F_{AB}$ vanishes (whichever occurs first). In the case where
$F_A(n_A^*)=0$ [while either $F_{AB}(n_A^*)<0$ or Eq. (\ref{eq:phasep1}) cannot be solved], lowering $n$ further leads to phase separation into phase A (with $n_A=n_A^*$)
and the EI phase; the EI fraction expands until phase A disappears at half-filling ($n=1$) and
the system turns into a uniform EI. At this point, the value of the chemical potential
suffers a negative jump (within the EI spectral gap), and reducing $n$ further leads to a phase
separation into EI and a single-band phase with empty narrow band. 

If, on the other hand,
$F_{AB}(n_A^*)$ vanishes [while $F_A(n_A^*)$ is still negative], this is followed by phase separation into phase $A$ (with $n_A=n_A^*$)
and phase B, with the density $n_B^*$ determined by $\mu_B(n_B^*)=\mu_A(n_A^*)$. The fraction
of phase $B$ then increases until we reach the point $n=n_B^*$, beyond which the system remains in the uniform
phase-B state with decreasing $n_B=n$. In the case where
$n_B^*<1/2<n_A^*$, the EI state does not arise: at
half-filling, its energy $E_{EI}$ is larger than the energy of an appropriate A-B phase mixture,
\begin{equation}
  \left[(1-n_B^*)E_A(n_A^*)+(n_A^*-1) E_B(n_B^*)\right]/(n_A^*-n_B^*)\,.
\end{equation}

\begin{figure}
\includegraphics{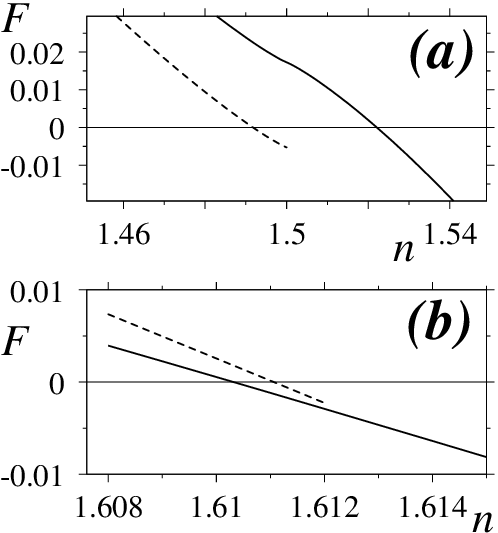}
\caption{\label{fig:F}
  Values of $F_A(n)$ (solid) and $F_{AB}(n)$ (dashed) for $U=2$, $E_d=0.4$, $t^\prime=-0.15$ {\it (a)} and for $U=0.5$, $E_d=0.4$, $t^\prime=-0.015$ {\it (b)}.}
\end{figure}

These two situations are exemplified by Fig. \ref{fig:F}. For the values of
parameters used in Fig. \ref{fig:F} (a), Eq. (\ref{eq:gap}) is satisfied
at $n<1.68$, and $F_A(n)$ changes sign at $n_A^*\approx 1.522$, signalling
phase separation into a half-filled EI and the single-band phase A with $n=n_A^*>1$ (phase separation of the type PS1). On the other hand,
the value of $F_{AB}$, which would describe phase separation into two different
single-band phases, changes sign at $n\approx 1.492 < n_A*$, and this latter
scenario is therefore irrelevant (as the uniform phase A at this doping level
is already unstable with respect to the other type of phase separation).
In the case of Fig. \ref{fig:F} (b), Eq. (\ref{eq:phasep1}) for those
parameters values can be solved only in a very narrow doping range $1.608 < n < 1.612$ [cf. Fig. \ref{fig:sbsm} (b), where the overall negative slope of the dashed line in the
  centre of the figure is very small]. The value of $F_{AB}(n)$ changes sign
at $n_A^*\approx 1.611$, whereas $F_A(n)$ vanishes at $n\approx 1.610< n_A^*$
and the EI state is therefore irrelevant.
Phase separation is thus of the PS2 type, corresponding to a mixture of two single band phases with $n=n_A^*$ and $n=n_B^*\approx 0.756$. We note that the
 precise boundary between the two scenarios depends also on the value of $t^\prime$, as increasing the latter tends to tilt the balance in favour of PS1.

While generally the PS2 phase separation may also involve SB2 and semimetal
phases, the specific case above corresponds to a phase equilibrium between
two different SB1 phases (with filled and empty narrow band). Apparently, this
can be
identified as a situation found in numerical investigations of doped FKM and
asymmetric Hubbard model, whereby the narrow-band electrons tend to clump
together in a part of the system (see, {\it e.g.}, Refs. \onlinecite{Farkasovsky08_2,Maska05})

We are now finally in a position to discuss the phase diagrams emerging
from our study.

\section{THE PHASE DIAGRAMS}
\label{sec:phadiag}

\begin{figure*}
\includegraphics{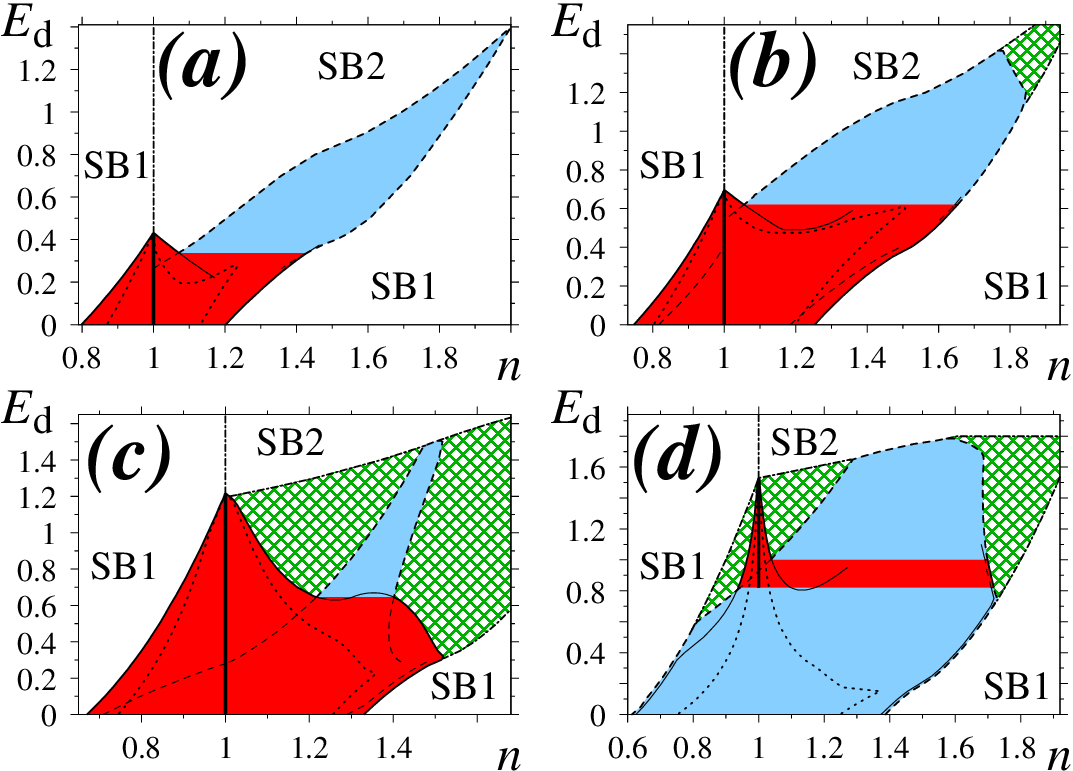}
\caption{\label{fig:phadiag}
(colour online)  Calculated phase diagrams for the 2D EFKM with $U=4$ and $t^\prime=-0.3$ {\it (a)},
  $U=2$ and $t^\prime =-0.15$ {\it (b)}, $U=1$ and $t^\prime=-0.1$ {\it (c)}, $U=0.5$
  and $t^\prime=-0.015$ {\it (d)}. The shaded regions
  correspond to phase separation: red (darker grey in the black and white
  version) to PS1, whereby one of the component phases is the half-filled EI state,
  and light blue (lighter grey) to PS2. Solid and dashed lines correspond to
  solutions of Eqs. (\ref{eq:gap}--\ref{eq:F}) and
  (\ref{eq:phasep1}),(\ref{eq:FAB}) [PS1 and PS2 conditions, respectively]; bold vertical solid lines at $n=1$ denote a stable single-phase EI. Dotted lines are the boundaries of the region around $n=1$, where the EM solution exists
  and has the lowest energy among the uniform single-phase states.
  Outside the phase separation region, the dashed-dotted lines show
  boundaries between single-band phases SB1 and SB2 (with the Fermi level within the broad or the narrow band, respectively) and semimetallic state (hatched).} 
\end{figure*}

In Fig. \ref{fig:phadiag} we present the phase diagrams corresponding to
different values of $U$ and $t^\prime$. The values of the latter were chosen
to be sufficiently large, so that the EI state at half-filling is expected to be stable with respect to low-lying collective excitations \cite{prb2012}.
On the other hand, the bare bandwidth of the narrow band remains several times
smaller than $U$ in all cases. The phase diagrams are of course symmetric under
the transformation $E_d \rightarrow -E_d$, $n \rightarrow 2-n$.

Phase separation is confined to the shaded regions of the phase diagrams.
The red region corresponds to PS1, whereby one of the two component phases is
the half-filled EI. As explained in the previous section, the state of
the system for any $n>1$ ($n<1$) corresponds to an appropriate mixture of the
EI phase and the uniform
phase which borders the
PS1 region above (below) half-filling at the same  $E_d$, with the value of
density $n$ at the border. Likewise, the state of the system within the PS2
(light-blue) region is the mixture of the two phases which border the phase
separation region from right and left at a given $E_d$.

The stable EI phase
is represented by a thick vertical  solid line at $n=1$. The upper
tip of this line corresponds to EI formation in a semiconductor, whereas in the
lower part excitonic pairing occurs in a metallic ``parent'' state. In order
to locate the crossover between these two regimes, we have to formally set
the off-diagonal average $\Delta$ to zero and require that the conduction and
valence Hartree bands touch:
\begin{equation}
  Un_d+2=E_d+Un_c-2|t^\prime|\,.
\label{eq:cross}
\end{equation}
While in the weakly-interacting case of Fig. \ref{fig:phadiag} {\it d} this
equality is satisfied (within our numerical accuracy) at the upper point of
the EI line, $E_d \approx 1.525$,
in case of the strong interaction (Fig. \ref{fig:phadiag} {\it a}) this takes
place at $E_d \approx 0.24$, near the middle of the EI line which extends up to
$E_d \approx 0.43$. In the intermediate cases of Fig. \ref{fig:phadiag} {\it b} and {\it c}, Eq. (\ref{eq:cross}) is satisfied, respectively, at $E_d \approx 0.56$ and $E_d \approx 1.205$, while the upper end of EI line corresponds to $E_d\approx 0.695$ and $E_d \approx 1.215$. 

The observation that
{\it the PS1 area (and associated excitonic behaviour) can extend relatively far
  away from the $n=1$ line} is among the main results of the present study.
We see that the PS1 region is most prominent at the intermediate values of $U$.
At large $U$ (Fig. \ref{fig:phadiag} {\it a}) it is suppressed due to large
Hartree contribution to particle energy in the EI state, whereas at small $U$
(Fig. \ref{fig:phadiag} {\it d}) the energy gain associated with the excitonic
pairing becomes marginal.

Bold solid (dashed) lines at the borders of the PS1 (PS2) regions correspond to
solutions to Eqs. (\ref{eq:gap}--\ref{eq:F}) [for PS2, Eqs. (\ref{eq:phasep1})
and (\ref{eq:FAB})]. Lighter lines correspond to continuations of these
solutions within the phase separation regions; while these do not correspond
to any transitions (see above, Sec. \ref{sec:phasep}), they help visualise
the overall structure of the phase diagram. We note that all four diagrams,
which sweep different parameter regimes from strong to weak interaction,
are remarkably similar in this regard. Therefore there is no doubt that
results for other parameter values
would be qualitatively similar.

The parameter space outside the PS1 and PS2 regions is occupied, for the
most part, by the single-band phases SB1 (with partially-filled broad band)
and SB2 (whereby the chemical potential lies within the narrow band).
In Fig. \ref{fig:phadiag}, the boundaries of the corresponding regions are
shown with dashed-dotted lines. While the boundary at half-filling is
continuous\cite{bandins} (with increasing density, the broad band is filled at $n=1$ and
subsequently the filling of the narrow band begins), no direct continuous
transition between SB1 (with filled narrow band) and SB2 phases is possible
in the $2>n>1$ region. In the large-$U$ case (see Fig. \ref{fig:phadiag}
{\it a}), the intervening area is entirely taken over by phase separation, PS2.
In a completely filled system at $n=2$, the difference between SB1 and SB2
phases disappears, hence the boundaries of these two regions must meet (which
is indeed the case for all parameter values used in Fig. \ref{fig:phadiag}).
This occurs when the upper edges of the two filled Hartree bands coincide,
\[U+2=U+E_d+2|t^\prime|\,,\]
or $E_d=2-2|t^\prime|$. The other end of the SB2-region lower boundary
(more precisely, its continuation within the PS1 region)
crosses the EI line at half-filling in the general area of a crossover between
semiconducting and metallic excitonic pairing [see Eq. (\ref{eq:cross})],
as can be expected.

Another non-excitonic phase, a semimetal with two partially-filled bands,
is strongly disfavoured at large $U$ due to a large Hartree contribution
to its energy. We see that as the value of $U$ decreases, areas of semimetallic
phase (hatched) emerge in the intervening region between SB1 and SB2.
For smaller-$U$ cases shown in Fig. \ref{fig:phadiag} {\it c,d}, we see that
at certain values of $E_d$ the phase separation (PS2) may occur also between
semimetallic phases with different values of density. This is because uniform
semimetallic solutions typically possess negative compressibility in a certain
range of values of $n$ [see above, Eq. (\ref{eq:smcompre})]; in
addition,
with varying density a discontinuous
switching between different semimetallic solutions (see Sec. \ref{sec:mf})
can occur,
also entailing phase separation.

Within our model, Eq. (\ref{eq:FKM}), the excitonic metal (EM) phase with
$n\neq 1$ has a negative compressibility (see Sec. \ref{sec:uniform} and Appendix) and does
not appear on the phase diagram. Dotted lines
within the phase separation regions in Fig. \ref{fig:phadiag} show the areas
around half-filling where the EM phase has the lowest
energy among the uniform single-phase solutions. For weaker $U$ of
Fig. \ref{fig:phadiag} {\it c,d} the latter  is always the case whenever the
EM solution exists. For $U=2$, (see Fig. \ref{fig:phadiag} {\it c}), there
is a narrow region immediately above the upper (concave) boundary
of the EM area at $n>1$, where the EM solution is present, yet has a higher
energy than a single-band one. Within and around this region we sometimes
encounter a situation where several EM solutions are present; the
lowest-energy one has a negative compressibility. Finally, at $U=4$ this region extends upwards and to the right far into the larger-$E_d$,$n$ range, although in some
cases the EM solution there may be
spurious (corresponding to a local energy {\it maximum}). Elsewhere, the
boundary of the EM region corresponds to a continuous transition ($\Delta \rightarrow 0$) into the lowest-energy uniform non-excitonic phase.
We also note that
the EM area significantly expands whenever the absolute value of $t^\prime$ is
decreased. We will continue our discussion of the EM phase in the following
section. 

We recall that  our selection of single phase states and component phases
for phase separation includes uniform mean-field solutions only. We expect
that this affects the validity of our results primarily at commensurate
fillings, especially at $n=1$ beyond the region where the uniform EI phase is
stable (which includes the charge/orbital ordering at small $|E_d|$, see Refs. \onlinecite{Batista04,Farkasovsky08,Czycholl08}).
Possible instability of a uniform semimetal at a fractional filling with
respect to charge/orbital modulation is
another issue which falls beyond the scope of this work.  Finally, when the
narrow-band hopping $t^\prime$ is decreased below the critical value required
to stabilise the uniform EI, the EI state at $n=1$ acquires spatial
modulation of both charge density and excitonic
correlations\cite{Farkasovsky08}. We expect
that this should not significantly affect the behaviour in the doped regime at a small but finite $t^\prime$:
within the overall picture described above, the modulated EI would take
place of the
uniform EI as a phase component in the PS1 region;
apart from a quantitative change, the overall structure of the phase diagram
presumably remains unaffected.

\section{DISCUSSION}
\label{sec:conclu}

The possibility of excitonic condensation, resulting in a formation of EI
state in a metallic or semiconducting compound at a low temperature, attracts
broad experimental
and theoretical effort (see Ref. \onlinecite{Kaneko2025} for a contemporary review). This includes theoretical
studies of doped systems, with applications to hexaborides\cite{Balents2000,Veillette2001,Ichinomia2001,Bascones2002}, twisted bilayer
graphene\cite{Ghorai2023}, and other compounds\cite{Bi2021}. These broader-band systems are generally treated within the
low-energy, long-wavelength approach, which in its original
form involves re-structuring of the spectrum in the immediate vicinity of
the (nested) Fermi surfaces, and lowering the net energy slightly by opening
a narrow
gap at the Fermi level\cite{Keldysh}. When doped, the proximity of the EI
state affects the properties of the system only as long as the chemical
potential lies close to the (possibly smeared) excitonic gap, corresponding to
a narrow range of doping values (cf. Appendix). Within this range, a rich physical picture
emerges once
additional degrees of freedom (most notably spin) and features like
imperfect nesting\cite{Balents2000,Veillette2001,Ghorai2023} or presence of
impurities\cite{Ichinomia2001} are taken into account. Findings
include, {\it inter alia},
ferromagnetism\cite{Volkov1975,Zhitomirsky1999,Balents2000,Veillette2001,Ichinomia2001,Bascones2002},
excitonic metal behaviour\cite{Veillette2001,Bi2021,Ghorai2023,Zhitomirsky1999},
and phase separation\cite{Balents2000,Veillette2001,Bascones2002}.

However, to the best of our knowledge
the available theoretical literature on the extended Falicov -- Kimball model
away from half-filling does
not address the issue of excitonic correlations. It should be emphasised that
EFKM corresponds  to a rather different realisation of the EI, whereby the
gap is
comparable or larger than the width of the narrow band\cite{precise}, and the issue of
Fermi surface nesting is no longer relevant. The excitonic correlations in
the EFKM are inherently short-wavelength,
hence the entire spectrum throughout the Brillouin zone is  modified.
The EI gap itself is sufficiently broad to allow for an equilibrium with a
conducting phase in a wide range of carrier densities [we recall that the
chemical potential of this second phase, which depends on density,  must be located
within the EI gap, see Eq. (\ref{eq:gap}).]

Indeed, in Sec. \ref{sec:phadiag}
above we saw that under the right conditions, whenever
the EI state is stabilised at half-filling, it persists in a doped system as a
component phase in a phase-separated state in a relatively broad range of
carrier densities. In such a state, the electrical
current would be carried by the other phase component only, and percolative
transport 
behaviour is anticipated. Since the chemical potential lies within the EI
energy gap, the {\it average} density of states will show a broad depression
around the Fermi level. There is a number of intriguing physical issues,
including the Andreev-like scattering of carriers by the borders of the EI
regions\cite{Wang} and the overall evolution of the system with increasing temperature (cf. Ref. \onlinecite{prb2020}).

There is also another possibility, which might turn out to be relevant
for actual physical systems. These always include the long-range Coulomb
interaction, which dictates that single-phase areas in a phase-separated
system cannot grow beyond a certain size; both the Coulomb contribution and
the surface tension of the interphase boundaries\cite{boundaries} are increasing the
energy of the phase-separated state with respect to that of a homogeneous one.
Since the energy difference between phase-separated state and competing
single-phase states is typically rather small, this may destroy the phase
separation/inhomogeneity and stabilise the homogeneous behaviour. On the other
hand,
the {\it relative} energies of various uniform single-phase solutions at a
given value of carrier density remain unaffected.

In Sec. \ref{sec:phadiag} we saw that there is a sizeable region on a phase
diagram, where the excitonic metal phase has the lowest energy among the
uniform homogeneous states (see the dotted lines in Fig. \ref{fig:phadiag}).
The fact that {\it in the absence of the long range
  Coulomb interaction} the EM state was found to have a negative compressibility
(Sec. \ref{sec:uniform}) obviously has little bearing on the situation when the
Coulomb interaction is present. We therefore suggest that including the Coulomb
interaction might stabilise the homogeneous EM phase in a doped system.

Should this be the case, the density of states at the Fermi level will be
{\it increased}, and a broad energy gap will open above or below the
chemical potential. The (hybridised) carriers at the Fermi level will have
predominantly narrow-band character (see Fig. \ref{fig:dos}), which will
in turn affect the transport properties.
  
Either way, we expect that EI behaviour, which is conventionally associated
with  EFKM at half-filling, may affect the properties of a doped system
over a broad range of carrier densities. Therefore, whenever experimental
results suggest that the EI or EM behaviour persists beyond one or two
percent doping, this may imply that the EFKM-like picture of strong
short-range correlations is relevant. In particular, this might be the case for
1$T$-TiSe$_2$ (where the charge density wave, broadly attributed to excitonic
pairing, is cut off by a superconducting state at 4\%-6\% copper intercalation\cite{Morosan2006}), and likely also for Ta$_2$NiSe$_5$ (see Refs. \onlinecite{Chen2020,Song2023}). 

~

\acknowledgements

The author takes pleasure in thanking R. Berkovits   
for discussions.
This work was supported by the Israeli Absorption Ministry. 

\appendix
\section{
  Compressibility of the excitonic metal phase}

At the phenomenological level, the origin of the negative compressibility in
the EM phase away from half-filling is clear from Fig. \ref{fig:EM}. Both
above and below the half-filling the EM eventually merges with another phase,
and the difference between the values of $\mu$ at these two points, whether
negative (as in Fig. \ref{fig:EM} {\it b}) or positive
(Fig. \ref{fig:EM} {\it a}), is relatively small. However, the dependence of
$\mu(n)$ in the intervening EM phase must accommodate a large {\it positive} jump at $n=1$,
equal to the width of the EI energy gap. It is therefore quite natural that
away from this jump the value of $\mu$ decreases with $n$.

In order to analyse the underlying mechanism in more detail, we will turn to
the $t^\prime \rightarrow 0$ case, ignoring the other (collective-excitation)
instability\cite{prb2012} which arises there.
This is expedient because  the negative compressibility of EM
persists for $t^\prime \rightarrow 0$ as well and clearly has the same nature as for $t^\prime\neq 0$, while on
the other hand the $t^\prime = 0$ case is much more transparent analytically. 

At $t^\prime=0$, the quasiparticle energies
$\epsilon^{1,2}_{\vec{k}}=\epsilon_{1,2}(\epsilon_{\vec{k}})$, Eqs. (\ref{eq:epsilon12}--\ref{eq:xi}), are 
monotonously increasing functions of the tight-binding energy
$\epsilon_{\vec{k}}$. Thus for $n<1$ ($n>1$) the quasiparticle states in the
lower (upper) band are occupied for $\epsilon_{\vec{k}}<\epsilon_0$, and the
chemical potential $\mu$ is given by $\epsilon_1(\epsilon_0)$
[$\epsilon_2(\epsilon_0)$ for $n>1$]. When the carrier density is increased
by a
small $\delta n$, the chemical potential change
$\delta \mu=\delta_{\rm r} \mu+\delta_{\rm c} \mu$ is obtained
from Eq.(\ref{eq:epsilon12}) as a sum of the
``rigid'' bandstructure contribution,
\begin{equation}
  \delta_{\rm r}\mu=\frac{1}{2}\left( 1 \pm \frac{\xi_0}{\sqrt{\xi_0^2+4U^2\Delta^2}} \right) \delta\epsilon_0
  \label{eq:deltarn}
\end{equation}
    [with $\delta\epsilon_0=\delta n/ \nu(\epsilon_0)$ and
      $\xi_0=\xi(\epsilon_0)$, see Eq. (\ref{eq:xi})], which is always positive, and the interaction-induced
 ``correlation'' term,
\begin{equation}
  \delta_{\rm c} \mu= \frac{U\delta n}{2} \mp \frac{\xi_0U (\delta n_c-\delta n_d) + 4U^2 \Delta \delta \Delta}{2\sqrt{\xi_0^2+4U^2\Delta^2}}\,,
  \label{eq:deltacn}
  \end{equation}
which includes the effects of a self-consistent change of
$n_{c,d}$ and $\Delta$. The upper and lower signs in
Eqs. (\ref{eq:deltarn}--\ref{eq:deltacn}) correspond to $n<1$ and $n>1$
respectively. 

By varying $n$ in the mean field equations (\ref{eq:delta}--\ref{eq:ndt}) and
solving for
$\delta n_c- \delta n_d$ and $\delta \Delta$, we find that $\delta_{\rm c}\mu$ is
proportional to $\delta n$,
\begin{eqnarray}
  \delta_{\rm c}\mu&=&
\frac{(2 \xi_0 Y_0-2U \Delta Y_1-U\Delta)^2 +(4 U \Delta Y_0+ \xi_0 Y_1)^2}
     {16Y_0^2+4Y_1^2+2Y_1}\times \nonumber \\
     &&\times
\frac {2 U \delta n}{\xi_0^2+4U^2 \Delta^2}
     \,,
\label{eq:drmu}
\end{eqnarray}
where [cf. Ref. \onlinecite{prb2012}, Eqs.(B3-B4)]
\begin{equation}
  Y_0=\frac{1}{2} \int \frac{U^2 \xi \Delta \nu(\epsilon) d\epsilon}
  {(\xi^2+4U^2 \Delta^2)^{3/2}}\,,\,\,\,
Y_1=- \int \frac{2U^3 \Delta^2 \nu(\epsilon) d\epsilon}
{(\xi^2+4U^2 \Delta^2)^{3/2}}\,,
\label{eq:Y01}
\end{equation}
$\xi=\xi(\epsilon)$,
the integration limits are from $-d$ to $\epsilon_0$ for $n<1$ or
from $\epsilon_0$ to $d$ for $n>1$, and $d=2$ or $3$ is the dimensionality.

The denominator in Eq. (\ref{eq:drmu}) is {\it negative}, as proved in
Ref. \onlinecite{prb2012} (Appendix B). The proof, based on the required
stability
of the EI state at a fixed $n=1$ with respect to an ``external field'' $V_0$
[see Eq. (\ref{eq:Vpert})], can be directly generalised for the case of an
EM state away from half-filling.

The quantity $\delta_{\rm c}\mu/\delta n$ is therefore negative, and gives a
negative contribution to $\partial \mu/\partial n$. This has a clear physical
meaning: while in the absence of correlations the compressibility would have
been  given by $\delta_{\rm r}\mu/\delta n >0$, in reality the self-consistent
mean field band structure
changes with  $n$ in such a way that the net energy is minimised. The
associated energy reduction is of second order in $\delta n$, which yields a
negative term in the compressibility.

The EM phase is found in the region where the quasiparticle density of
states at the Fermi level is strongly increased due to hybridisation,
meaning $\xi_0 < 0$
for $n<1$, $\xi_0 >0$ for $n>1$ ({\it i.e.}, the Fermi level lies between the
crossing, $\xi(\epsilon)=0$, of the Hartree $\Delta \rightarrow 0$ bands and
the EM energy gap).
In the $U\Delta\ll |\xi_0|$ limit, the term  $\delta_{\rm r}\mu/\delta n$ then
vanishes, whereas $Y_1$ and hence $\delta_{\rm c}\mu/\delta n$ remain finite.
Similarly in the $U\Delta\gg 1$ limit, $\delta_{\rm r}\mu/\delta n$ remains
finite whereas $\delta_{\rm c}\mu/\delta n \propto -U\Delta$.
Both limits are formal since they can not arise as mean-field solutions;
the real $\Delta \rightarrow 0$ regime at $t^\prime =0$ corresponds to either
$n_d \rightarrow 0$ or $n_d \rightarrow 1$. Yet the fact that the negative
$\delta_{\rm c}\mu$ dominates in both cases, along with the results of our
numerical investigation of actual mean field solutions for the 2D case,
suggests that this is
indeed the generic behaviour, both in $t^\prime=0$ and $t^\prime \neq 0$ cases.

There is however one important exception, corresponding to the 3D case just
above or
just below half-filling. In the limit $n \rightarrow 1$, owing to vanishing
$\nu(\epsilon)$ at the band edge, the dominant term is
$\delta_{\rm r}\mu \propto (\delta n)^{2/3}$. This results in a {\it positive}
compressibility, presumably within an immediate vicinity of half-filling\cite{1D}.

Based on the above discussion of EM compressibility in an oversimplified
model, we can formulate tentative conclusions which perhaps are not
restricted to the EFKM as described by  Eq.(\ref{eq:FKM}): (i) $\delta \mu$ contains two
terms, corresponding to filling a rigid hybridised bandstructure and
to correlation
effects. The correlations  term is
negative, always linear in $\delta n$ (because it comes from a
variation of Fermi-sea integrals) and typically dominant, resulting in a
negative $\partial \mu/\partial n$ (ii) In three dimensions, whenever the Fermi
level approaches a band edge or a Van Hove anomaly, the compressibility
turns positive, presumably within a narrow range of doping values. Thus the EM
{\it may} be thermodynamically stable in this region, provided that there are
no other instabilities. In this
regard, we note that the purported excitonic ferromagnetism in hexaborides
is limited to small values of doping, although the
difference in both physical systems and theoretical approaches does not
allow for a direct analogy (see discussion in Sec. \ref{sec:conclu}).


\begin{thebibliography}{999}
\bibitem{bandins} The vertical dashed-dotted lines at $n=1$ in
  Fig. \ref{fig:phadiag} correspond to a band insulator; chemical potential in
  all the relevant phases at $n>1$ lies above its gap, $\mu>E_d+U-2|t^\prime|$,
  hence this insulator (unlike EI) cannot become a component in an equilibrium
  phase mixture.
\bibitem{precise} More precisely, the gap is larger than the narrow-band
  Fermi sea depth (energy difference between the Fermi level and the
  narrow-band edge in the unhybridised ``parent'' bandstructure with the
  same value of $n_d$).
\bibitem{1D} Note that when a finite  $t^\prime<0$ is included in
  Eq. (\ref{eq:FKM}), the extrema of the hybridised bands above and/or below
  the gap shift away from $\epsilon_{\vec{k}}=\pm 2$, and the density of states at the band edge acquires
  one-dimensional character (in both 2D and 3D cases). This occurs once
  $|t^\prime|$ exceeds a certain value (which is numerically small),
  and is an artefact of our tight-binding model, Eq. (\ref{eq:FKM}). In this
  situation, the $\delta_{\rm r}\mu$ term in the vicinity of $n=1$ becomes
  negligible, due to $\delta \epsilon_0 \propto (\delta n)^2$.
\end{thebibliography}
\end{document}